\documentclass[aps,prb,twocolumn,groupedaddress]{revtex4}
\usepackage{graphicx}
\usepackage{amsmath}

\begin{document}

\title{Reconstructing Three-dimensional Helical Structure With an X-Ray Free Electron Laser}
\author{M. Uddin}
\email[e-mail address:\,\,\,]{miraj.uddin@uwc.edu}
\affiliation{Department of Physics, University of Wisconsin-Milwaukee,
Milwaukee, WI 53211} 

\date{\today}

\begin{abstract}

Recovery of three-dimensional structure from single particle X-ray scattering of completely randomly oriented diffraction patterns as predicted few decades back has been real due to the advent of the new emerging X-ray Free Electron Laser (XFEL) technology. As the world's first XFEL is in operation starting from June 2009 at SLAC National Lab at Stanford, the very first few experiments being conducted on larger objects such as viruses. Many of the important structures of nature such as helical viruses or deoxyribonucleic acids (DNA) consist of helical repetition of biological subunits. Hence development of method for reconstructing helical structure from collected XFEL data has been a top priority research. In this work we have developed a method for solving helical structure such as TMV (tobacco mosaic virus) from a set of randomly oriented simulated diffraction patterns exploiting symmetry and Fourier space constraint of the diffraction volume. 

\end{abstract}

\maketitle
\section{Introduction}
In 1977 Kam \cite{Kam1977}   pointed out that correlated fluctuations in intensity in x-ray scattering of non-oriented identical particles contain useful structural information regarding the particle itself. In fluctuation scattering experiment, radiation must be recorded on time scale shorter than the rotational diffusion time \cite{Elser2011}. Since the XFELs are in operation in USA and elsewhere, the development of useful technique and algorithm for structure determination from collected diffraction patterns of random orientations from diffract and destroy experiment is very important not only from technological point of view but also for finding clues for diseases and designing drug to treat them. So far structure determination of helical biological structure such as TMV\cite{Namba1986} or DNA \cite{Watson1953} has been primarily done by fiber diffraction experiment where the helical structures are aligned along their body long axis which is tedious because of the entropic tendency of the molecule. In this work we are reporting a full 3D recovery of TMV up to three repeating unit from a set of simulated diffraction patterns of randomly oriented TMV helices (atomic coordinate of biological assembly of TMV  deposited in protein data bank as 2tmv ) in XFEL diffract and destroy experiment. \\ 
\section{Background}
Scattered intensity distribution $I$  over 3D molecular reciprocal space in spherical coordinate may be expressed in spherical harmonic expansion as shown by Saldin {\it{et. al.}} \cite{Saldin2009}
\begin{equation} 
I(q, \theta, \varphi)=\sum_{LM} I_{LM}(q) Y_{LM}(\theta, \varphi)
\label{int}
\end{equation}
where $Y_{LM}(\theta, \varphi)$ is a spherical harmonic ($L$ and $M$ are usual angular momentum quantum numbers).
In principle a full three dimensional (3D) structure of a biological object may be reconstructed provided the above expansion coefficients can be recovered from the collected set of two dimensional (2D) diffraction patterns of random orientations. 

Each diffraction pattern of random orientation represents a section through the three dimensional (3D) reciprocal space of the molecule. The intensities on the curved section of the Ewald sphere\cite{Saldin2009} may be labelled by magnitude q of the scattering vector and an azimuthal angle $\varphi$. The angular cross correlation function between intensities of two different resolution rings $q$ and $q'$ averaged over a set of diffraction patterns defined as 

\begin{equation}
C_{2}(q,q',\Delta \varphi)=\frac{1}{N_{sj}} \sum_{sj} \sum_{n} I_{sj} (q, \varphi_{n}) I_{sj} (q', \varphi_{n} + \Delta \varphi)
\label{twoptcorr}
\end{equation}
Where $I_{sj} (q, \varphi_{n})$ is the intensity of a pixel on the $sj$-th diffraction pattern and $N_{sj}$ is the number of diffraction patterns of random orientations consisting the primary collected data set. 
Note that the orientational averaging of the diffraction patterns in Eq.(\ref{twoptcorr}) is a reasonable assumption since for a large number of diffraction pattern all orientations are equally likely \cite{Pande2014} in SO(3) space suggesting that the left hand side of Eq.(\ref{twoptcorr}) will be independent of the value $\varphi$ chosen on the right hand side.

\begin{figure}[htbp]
\centering\includegraphics[width=5.5cm]{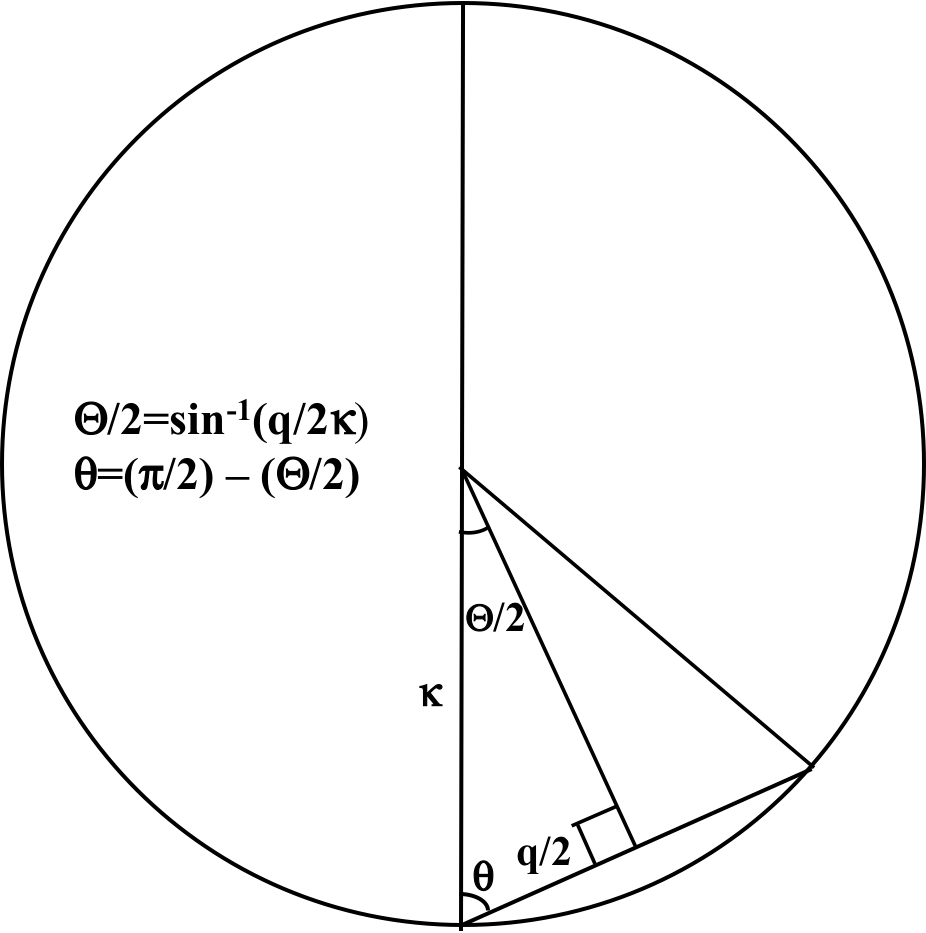}
\caption{Schematic of the diffraction geometry ($\kappa$ is the magnitude of the incident wave vector, $\Theta$ is scattering angle) shows the relationship between the polar angle $\theta(q)$ and the scattering angle $\Theta(q)$ as a function of the magnitude of the scattering vector $q$. Note that for flat Ewald sphere $\theta=\pi/2$.}
\label{SpEwald}
\end{figure}

Intensity distribution on the $s1$ diffraction pattern may be expressed as

\begin{equation}
I_{(s1)}(q, \varphi)=\sum_{LMM'} D^{(s1)}_{LMM'} I_{LM}(q) Y_{LM'}(\theta (q), \varphi)
\label{Wigmatrix}
\end{equation}

Similarly, intensity distribution on the $s2$ diffraction pattern may be expressed as

\begin{equation}
I_{(s2)}(q', \varphi')=\sum_{L'M''M'''} D^{(s2)}_{L'M''M'''} I_{L'M''}(q') Y_{L'M'''}(\theta' (q'), \varphi')
\label{Wigmatrix2}
\end{equation}

Substitution of Eq.(\ref{Wigmatrix}) and Eq.(\ref{Wigmatrix2}) into Eq.(\ref{twoptcorr}) and the orthogonal property of the Wigner D matrices  in SO(3) (Eq.(\ref{DOrtho})) allows the addition of two spherical harmonics (Eq.(\ref{AddSpHarmonics}))  in Eq.(\ref{twoptcorr}).

\begin{equation}
\frac{1}{N_{sj}} \sum_{sj} D^{(s1)}_{LMM'} D^{(s2)}_{L'M''M'''}=\frac{1}{2L+1} \delta_{LL'} \delta_{MM''} \delta_{M'M'''}
\label{DOrtho}
\end{equation}

\begin{equation}
\sum_{M} Y^{*}_{LM} (\theta (q), \varphi) Y_{LM} (\theta' (q'), \varphi') =\frac{2L+1}{4 \pi} P_{L} [\cos \gamma] 
\label{AddSpHarmonics}
\end{equation}

\begin{figure}[htbp]
\centering\includegraphics[width=8cm]{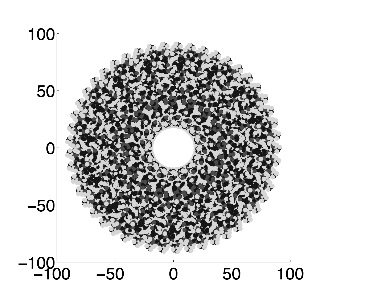}
\caption{Cross-sectional view of a single unit of TMV shows the $49$-unit protein subunit protrudes out along the helix. The axes are in real-space coordinates in units of $\AA$.}
\label{crosssection}
\end{figure}

Above substitution finally lead to the following simplification of the intensity correlation function \cite{Saldin2009}  as shown in Eq.(\ref{MatInv})

\begin{equation}
C_{2}(q,q',\Delta \varphi) = \sum_{L}  B_{L}(q,q')  \frac{1}{4 \pi} P_{L} [\cos \gamma] 
\label{MatInv}
\end{equation}

where
\begin{equation}
\cos \gamma=\cos \theta (q) \cos \theta' (q') + \sin \theta (q) \sin \theta' (q') \cos (\Delta \varphi) 
\label{Abbr1}
\end{equation}
and
\begin{equation}
B_{L}(q,q')=\sum_{M} I_{LM}(q) I_{LM}^{*} (q')
\label{Blq1}
\end{equation}
\subsection{Matrix Inversion}

$B_{L}(q,q')$ can be obtained from $C_{2}(q,q',\Delta \varphi)$ Eq.(\ref{MatInv}) using a matrix inversion technique based on singular value decomposition (SVD) which is purely a quantity obtained from diffraction patterns. The singular value decomposition of a matrix $A(m \times n)$ where $m < n$ is the factorization of $A$ into the product of three matrices

\begin{equation}
A=U \Sigma V^{T}
\label{SVD01}
\end{equation}
where $U(m \times m)$ and $V(n \times n)$ are orthonormal matrices Eq.(\ref{SVD02}), Eq.(\ref{SVD03}) and $\Sigma(m \times n)$ is a  diagonal matrix having only nonnegative diagonal entries in ascending order such that $(\sigma_{1} \leq \sigma_{2} \leq ......\leq \sigma_{m} \leq 0)$.

\begin{equation}
UU^{T}=I
\label{SVD02}
\end{equation}

\begin{equation}
VV^{T}=I
\label{SVD03}
\end{equation}

The zero padding of $\Sigma(m \times n)$ matrix can be carved to compute the $[\Sigma'(m \times n)]^{-1}=\Sigma''(n \times m)$ matrix to finally obtain the $A^{-1}$ matrix by Eq.(\ref{SVD04})

\begin{equation}
A^{-1} = (V^{T})^{-1} \Sigma'' U^{-1}
\label{SVD04}
\end{equation}
Note that in correlation method a large data set consisting of huge (ideally infinite)  number
of diffraction patterns has been reduced into a compact data set consisting of quadratic functions of Fourier shell coefficients. 

However recovering $I_{LM}(q)$ coefficients from the known quadratic shell correlation term $B_{L}(q,q')$ is a formidable mathematical and computational challenge and development of such a method in principle would allow a general approach for solving structure using correlation method. While development of such a method is yet to be accomplished, biological structure having certain symmetry such as helical symmetry \cite{Poon2013} or icosahedral symmetry \cite{Saldin2011} may be solved using method based on primarily symmetry assistance technique \cite{Kirian2012}.

TMV consists of repeating unit of biological building block \cite{Klug1999} and each unit consists of $49$ protein sub-unit spanned along the three-turn (cross-sectional view of single repeating unit shown in Fig. (\ref{crosssection})) provides the clue that the $M$ values in Eq.(\ref{Blq1}) be $0$, $\pm 49$, $\pm 98$, etc. 

Since $M \le L$, if we limit $Lmax=48$, the only permitted value of M be zero. Hence Eq.  (\ref{int}) for the case of TMV may be modified as
\begin{equation}
I(q, \theta, \varphi)=\sum_{L} I_{L0}(q) Y_{L0}(\theta, \varphi)
\label{intM0}
\end{equation}
i.e.; up to $L_{max}=48$, the TMV diffraction volume may be reconstructed from $M=0$ component alone. Since the radius of TMV is about $\sim ~100$ $\AA$ and each $c$ repeat unit \cite{Millane1991} of TMV is $69$  $\AA$, the conventional wisdom for angular momentum definition \cite{Saldin2011}, $q_{max} \times R=Lmax$; permits a reconstruction of TMV repeating unit  up to a $q_{max} \simeq 0.5$ $\AA^{-1}$.

\begin{figure}[htbp]
\centering\includegraphics[width=9cm]{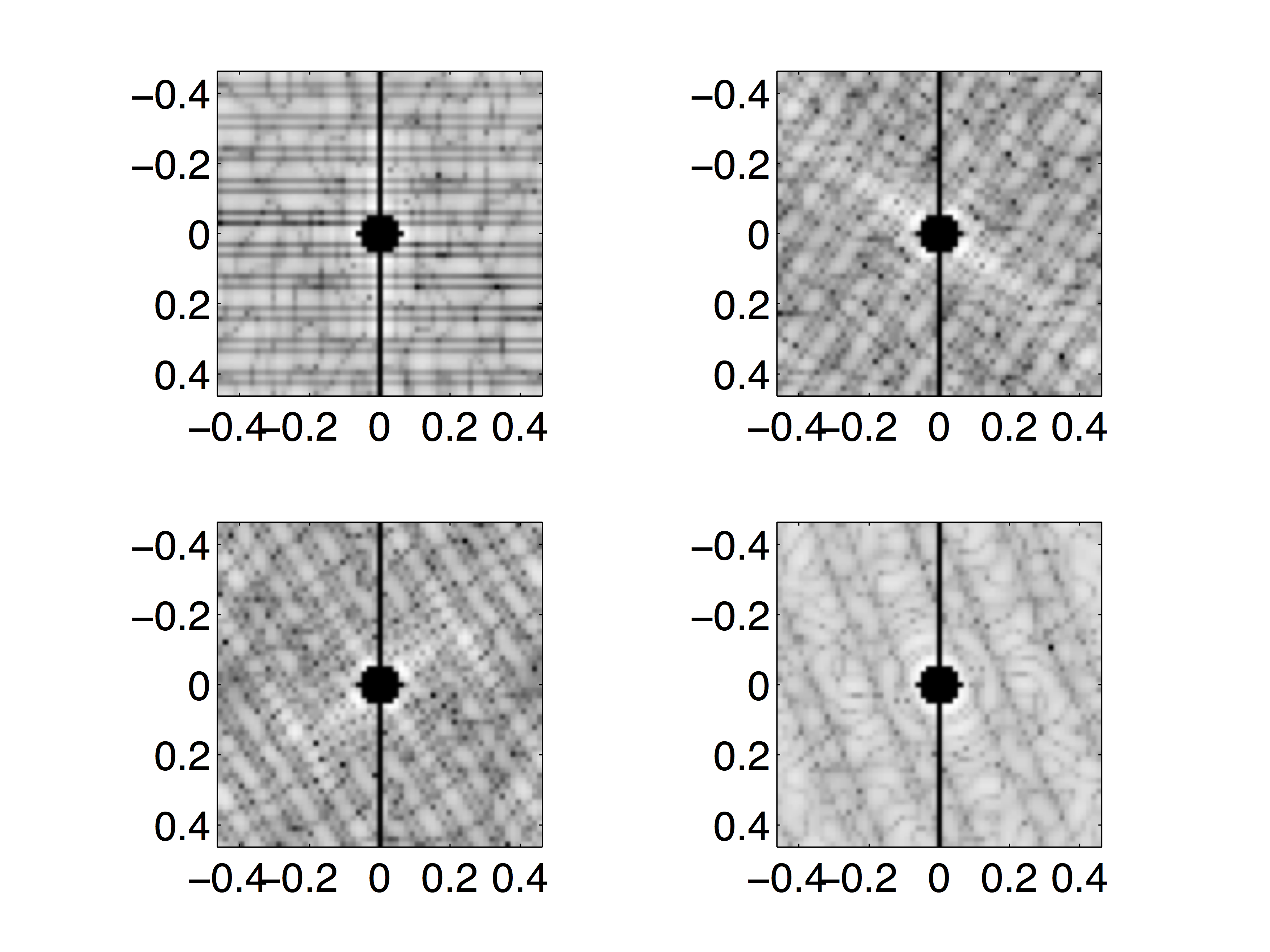}
\caption{Diffraction patterns for various random orientations of TMV. TMV orientations for which the X-ray incident beam is perpendicular to the body long axis (upper left),  $\sim \pm 45^{0}$ to the body long axis (upper right, lower left)  shows the layer line features are separated by 
$2 \pi /c$ where $c$ is the length of repeat unit of TMV which is $69$ $\AA$. Orientations for which the body long axis is somewhat parallel to the incident X-ray beam does not show $2 \pi /c$ splits of layer line features (lower right). 
}
\label{dps}
\end{figure} 

Eq.(\ref{Blq1}) holds the primary relation among various Fourier shells (including the diagonal term as well as the cross term) in quadratic form in the reciprocal space of the molecule. To recover $I_{L0}(q,q)$ coefficients from $B_{L}(q,q)$, a triple correlation function \cite{Poon2013}$^{,}$\cite{Kam1982} Eq.(\ref{threePtcorr}) and the associated three-point angular correlation has been introduced Eq.(\ref{Tlq1})

\begin{equation}
T_{L}(q)=\sum_{L_{1}L_{2}} G(L_{1}0, L_{2}0; L0) I_{L_{1}0} (q)  I_{L_{2}0} (q)  I_{L0} (q)
\label{Tlq1}
\end{equation}
where $G$ is a Gaunt Coefficient.

The ring triple correlation is related to experimental three point angular correlation $C_{3}$ as
\begin{equation}
C_{3}(q, \Delta \varphi)=\int T_{L} (q) P_{L}[\cos (\Delta \varphi)]
\label{threePtcorr}
\end{equation}

\begin{figure}[htbp]
\centering\includegraphics[width=8cm]{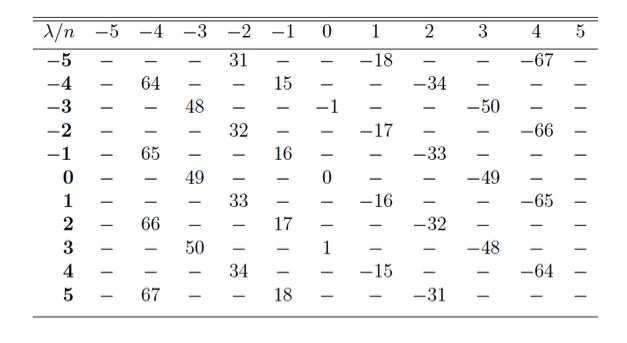}
\caption{Helix selection rule of TMV $49_{3}$ helix. Layer line index lambda showed along $Y$ in bold. We see that the allowed values of cylindrical harmonic $n-n'=49N$,  where $N=0, \pm 1, \pm2$, .. etc.
}
\label{SelRuleTab}
\end{figure}

To assist further why the M=0 approximation is a valid assumption for the reconstruction of helical bio-structure whose Fourier transform in three dimensional (3D) reciprocal space shows layer discs construction preferably expandable in reciprocal space cylindrical coordinate $(\mathcal{R}, \psi, \zeta)$ in terms of cylindrical harmonic \cite{Poon2013}$^{,}$ \cite{Cochran1952}$^{,}\cite{Saldin2010}$ as 
in Eq.(\ref{CylHarEx}).

\begin{equation}
I(\mathcal{R}, \psi, \zeta)=\sum_{n,n'} G_{n} (\mathcal{R}, \zeta_{\lambda}) G_{n'}^{\ast} (\mathcal{R}, \zeta_{\lambda'}) e^{i (n-n') \psi}
\label{CylHarEx}
\end{equation}
where $G_{n}(\mathcal{R}, \zeta_{\lambda})$ is a cylindrical harmonic ($\zeta_{\lambda} =  2 \pi \lambda /c$), as expressed in Eq.(\ref{Gharmonic}) associated with $\lambda$ -th layer line 
(permitted $n$ values for various layer line number are tabulated in Fig.(\ref{SelRuleTab}) .

\begin{equation}
G_{n}(\mathcal{R}, \zeta_{\lambda})=\sum_{k} i^n f_{k}   J_{n}(i \mathcal{R} r_{k}) e^{i [\zeta_{\lambda} Z_{k} - n \phi_{k}]}
\label{Gharmonic}
\end{equation}

where $J_{n}$ is a n-th order Bessel function and the $k$ summation runs over the real space cylindrical coordinate $(r, \phi, Z)$  of the atoms of the helical structure.

 \subsection{Helix Selection Rule}
It has been shown \cite{Poon2013}$^{,}$\cite{Cochran1952}$^{,}$\cite{Klug1958} that the helical repetition of $49$-unit protein subunit along the $3$-turn introduces a $49_{3}$ helical symmetry that essentially lead to a helix selection rule for TMV in Fourier space Fig.(\ref{SelRuleTab}) as follows
\begin{equation}
\lambda = 3n+49m
\label{selectionrule}
\end{equation}
where $\lambda$ is layer line number and $n$ is associated to the order of cylindrical harmonic expansion.  
\begin{figure}[htbp]
\centering\includegraphics[width=6.5cm]{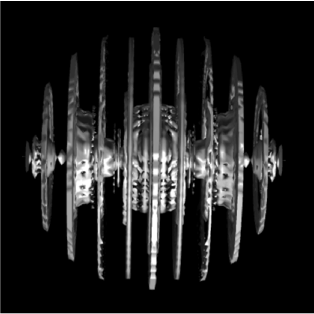}
\caption{Reconstructed diffraction volume for TMV $3$ unit shows the layer disc pattern spaced by $2 \pi / c$.
}
\label{diffvol}
\end{figure}

Due to helical arrangement of the protein subunits the layer line intensity of TMV expressed in terms of cylindrical harmonic Eq.(\ref{CylHarEx}). On the other hand, due to the random nature of the particle orientation subject to XFEL diffract and destroy experiment, the diffraction volume in correlation technique expressed conveniently in spherical geometry Eq.(\ref{int}). According to helix selection rule 
\begin{equation}
n-n'=M
\label{nnpM}
\end{equation}
and the only allowed values of $M$ for a $49_{3}$ helix are $49N$, where $N=0$, $\pm 49$, $\pm 98$, etc. Here the above claim concludes that M=0 term generate the first order harmonics for the expansion of the diffraction volume of TMV, thus reducing Eq.(\ref{Blq1}) as

\begin{equation}
B_{L}(q,q') = I_{L0} (q)  I_{L0} (q')
\label{Blq2}
\end{equation}

Eq.(\ref{Blq2}) allows the possibility of recovering the magnitude of diagonal Fourier shell coefficients by taking the square root of the $B_{L}(q,q)$ coefficients uncertain up to the signs of those discrete coefficients. More specifically
\begin{equation} 
I_{L0}(q,q) = \pm  |  \sqrt{B_{L} (q,q)} |
\label{sgnEqtn1}
\end{equation}
Structural information hidden in $B_{L}(q,q')$ obtained from a set of diffraction patterns of free electron laser or ultrabright synchrotron source results in unknown sign determination as in Eq.(\ref{sgnEqtn1}) in angular correlation method for certain biological samples having symmetry, as for example; the $49_{3}$ helical symmetry of TMV. 
Double phasing technique might be useful to start with a initial good guess of the signs of the above coefficients Eq.(\ref{sgnEqtn1}) and then to iteratively modify them based on reciprocal space constraints. The additional challenge for this approach involves the convergence of the reconstruction of the correct diffraction volume as well as the recovery of the real space object form the recovered diffraction volume by standard phasing technique.
The very recent work of Donatelli {\it{et. al.}} \cite{Donatelli2015} introduces a multitiered iterative phasing (MTIP) algorithm based on a series of derived projection operators to iteratively modify the specified real space constraints and to match the data to external observations. This technique provides a framework that would allow the extension of density modification techniques developed for crystallographic structure determination. Though this model does not require symmetry consideration, however the quality of the structure determination in this method is dependent on the accuracy of the data set as well as the amount of known priori information \cite{Donatelli2015}.
\begin{figure}[htbp]
\centering\includegraphics[width=7.25cm]{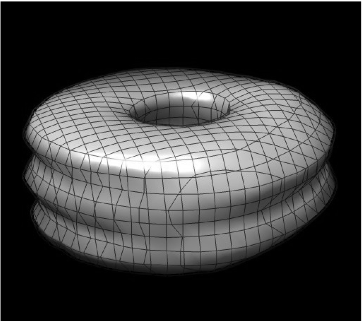}
\caption{Reconstructed single unit seen to fit the PDB mesh structure within the outer capsid.}
\label{ReconsOneUnit}
\end{figure}
\\
Another approach towards computing the scattering profiles from fluctuations X-ray scattering data based on three dimensional (3D) Zernike polynomial expansion model as introduced by Liu {\it{et. al.}} \cite{Liu2012}$^{,}$\cite{Liu2013}, demonstrates the feasibility of $ab$  $initio$ model reconstruction of nanoparticles from experimental data based on validation of several theoretical computation of representative molecules. Though this model does not assume any symmetry constraint, however it is computationally expensive \cite{Donatelli2015}.
\\
\\ 
On the other hand, known priori information may also be used to model the mathematical constraints in Fourier space thus reconstructing the diffraction volume with more accuracy and then to recover the structure using the standard phasing technique \cite{Oszlanyi2004}$^{,}\cite{Oszlanyi2005}$. The model introduced in this work uses some priori information regarding the structure of the molecule such as the internal diameter of TMV.
 \\
 \\
\section{Proposed Model}
The radial electronic charge density of TMV may be presented as an object constraint 1D step model (model 1) as following:

\begin{equation*}
\rho (r)= \begin{cases}
0 &\text{, $r < R_{1}$}\\
\rho_{constant} &\text{, $R_{1} \leq r < R_{2}$}
\end{cases}
\label{model1}
\end{equation*}
Where $R_{1}$ and $R_{2}$ are inner and outer radius of TMV. The inner core of TMV primarily composed of RNA (ribonucleic acid) surrounded by protein subunits. Since the electronic charge density distribution of RNA  is relatively higher than that of outer protein coat, the radial charge density variation of TMV may be modeled with a slowly varying 1D exponential term ($\xi \longrightarrow 0$)  as follows (model 2): 

\begin{equation*}
\rho (r)= \begin{cases}
0                       &\text{, $r < R_{1}$}\\
 \exp[- \xi r]   &\text{, $R_{1} \leq r < R_{2}$}
\end{cases}
\label{model2}
\end{equation*}
The scattered intensity is the squared modulus of complex amplitude; the Fourier Transform (FT) of the electronic charge density $\rho(\vec{r})$ of the real space object.

\begin{eqnarray}
I(\vec{q}) &=& | A(\vec{q}) |^{2}  \nonumber \\
               &=& | \int d^{3} \vec{r}  \rho (\vec{r}) e^{ i \vec{q}. \vec{r}} |^{2}\nonumber \\
               &\simeq&  | \int d^{3} \vec{r}  \exp[- \xi r] e^{ i \vec{q}. \vec{r}} |^{2}
\label{modelAddedEq}
\end{eqnarray}
Here $\vec{r}$ denotes the real space coordinate and $\vec{q}$ corresponding reciprocal space variable. When the object is rotated in SO(3) the radial variation in 3D real and reciprocal space is conjugated via the angular momentum.

Regarding the two rotational degrees of freedom in reciprocal space, the polar variation is intrinsically inserted in the reconstruction via $L$ quantum number and the azimuthal symmetry was imposed as described in Eq. (\ref{intM0}).

Assuming the radial variation in $\rho (r)$ is the prime focus of the model, Eq.(\ref{modelAddedEq}) may be treated as a 1D integral. With a limit $R_{2} \longrightarrow \infty$;  the 1D Fourier transform of the  slowly varying decaying exponential in $\rho$ can be written as:

\begin{equation}
T(k) = \int^{\infty}_{R_{1}} dr \exp[- \xi r] \exp[- i k r] 
\label{FFT1D1}
\end{equation}

With a limit $R_{1} \longrightarrow 0$, the Fourier transform integral can be evaluated in the following way:
\begin{eqnarray}
T(k)        &=& \int^{\infty}_{R_{1}} dr \exp[- \xi r] \exp[- i k r] \biggr \rvert _{R_{1} \longrightarrow 0} \nonumber \\
               &=& \frac{1}{ \xi + i k}  \nonumber \\
               &=&  \frac{\xi}{\xi^{2}+k^{2}} - \frac{ik}{\xi^{2}+k^{2}}
\label{FFT1D2}
\end{eqnarray}

With the limit $\xi \longrightarrow 0$ the real part of the Fourier transform Eq.(\ref{FFT1D2}) tend to a $\delta$ function 
in reciprocal space at $k=(\frac{2 \pi}{R_{1}})$ Eq.(\ref{deltaEq1}) .

\begin{equation} 
\frac{\zeta}{\xi^{2}+k^{2}}  \rightarrow  \pi \delta (k=\frac{2 \pi}{R_{1}})
\label{deltaEq1}
\end{equation}

And the imaginary part varies as $1/k$ Eq.(\ref{ImpartEq1})
\begin{equation} 
\frac{k}{\xi^{2}+k^{2}}  \rightarrow  \frac{1}{k}
\label{ImpartEq1}
\end{equation}

The existence of the $\delta$-like behavior in reciprocal space of the molecule (note that the contribution from the imaginary part is unimportant) due to real space charge density step boundary provide very important clue regarding the positivity of the diagonal shell correlation Eq.(\ref{sgnEqtn1})  of the resolution shell corresponding to $q=2 \pi/(N_{NS}  R_{1} )$ where $N_{NS}$ is the Nyquist oversampling rate \cite{Hayes1982} (for TMV $R_{1} \approx 19  \AA$). For two times intensity oversampled data $(N_{NS}=2)$, Eq.(\ref{sgnEqtn1}) can be written as

\begin{equation} 
I_{L0}(q,q=\frac{\pi}{R_{1}}) = +    \sqrt{B_{L} (q,q=\frac{\pi}{R_{1}})} 
\label{sgnEqtn02}
\end{equation}

Eq.(\ref{sgnEqtn02}) would be the decisive resolution shell whose signs propagates to all the corresponding shells. The justification of the above claim lies in the fact that the method described here is not limited to molecules having helical geometry; it has been verified for other geometry as well.

\begin{figure}[htbp]
\centering\includegraphics[width=8.0cm]{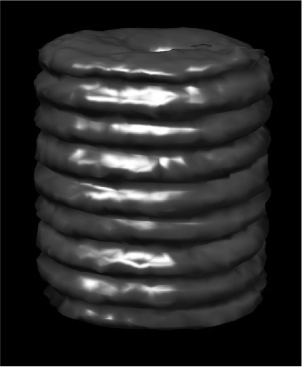}
\caption{Three-unit reconstruction of TMV. Reconstruction correctly recovers the helical turns as well as the central hole of TMV.
}
\label{ReconsThreeUnit}
\end{figure}

To be more specific; $I_{L0}(q,q= \pi /R_{1})$ is a decisive shell for the determination of other resolution shells via the quadratic cross correlation function $B_{L}(q,q')$ of Eq. (\ref{Blq2}) for reconstructing the three dimensional (3D) Fourier map of the molecule. 

\section{Results and Validation}
As an initial validation of the above method, roughly a set of one thousand simulated diffraction patterns of TMV was generated with an optimized code for simulating diffraction patterns which was the most computationally intensive part of the project.  For the simulation the generated diffraction patterns successfully account the contribution of the many repeating units from a single unit calculation via a shape transform factor. For some orientations the simulated diffraction patterns show the layer line features Fig. (\ref{dps}) separated by $2 \pi /c$ where $c$ is the length of single repeat unit of TMV.

From the set of simulated diffraction patterns $B_{L}(q, q')$ was recovered using the matrix inversion technique for non square matrices followed by a calculation of intensity cross-correlation using Eq.(\ref{twoptcorr}). Once a correct determination of the decisive correlation shell was obtained via the proposed model, determination of the remaining shells were obtained via Eq.(\ref{Blq2}).
Upon successful recovery of the shell correlation the reconstruction of the three dimensional (3D) diffraction volume was obtained by Eq. (\ref{intM0}).
Fig.(\ref{diffvol}) shows the layer disc reconstruction of the diffraction volume where the discs are separated by the characteristic reciprocal spacing $2 \pi /c$.
 
The real space reconstruction of the electron density of one unit and three unit  of TMV was obtained using standard phasing algorithm based on charge flipping method \cite{Oszlanyi2004}$^{,}\cite{Oszlanyi2005}$. The resolutions of both the reconstructions are $\sim 13 \AA$. 

To further assist the quality of the reconstruction, the reconstruction was superposed to the PDB reconstruction of TMV single repeating unit. 
Fig.(\ref{ReconsOneUnit}) shows the reconstruction closely fits the helical repetition as well as the central hole of TMV structure. The three unit reconstruction recovers the helical grooves as well as the central hole of TMV as shown in Fig.(\ref{ReconsThreeUnit}).

\section{Summary}
In this work  we have demonstrated the applicability of a technique that some known real space object constraints may be used to appropriately determine single resolution reference shell and the subsequent determination of other Fourier shells can be obtained via Fourier shell cross-correlation Eq. (\ref{Blq2}). 

We successfully applied the method for reconstructing single (and three) repeating unit of TMV from simulated XFEL diffraction patterns. Further improvement of the model might be useful for reconstructing object such as DNA double helix \cite{Watson1953} or objects having structures deviated partially from the above-mentioned case, as for example the coiled RNA core of TMV which protrudes out of the TMV inner capsid. For small deviation of the structure, we may consider the small variation in $B_{L}(q,q')$ of Eq.(\ref{Blq1}) to obtain
\begin{equation}
\delta B_{L} (q,q') = \sum_{M} [\delta I_{LM}(q) I^{\ast}_{LM}(q') + I_{LM}(q) \delta I^{\ast}_{LM}(q')]
\label{blqDelta01}
\end{equation}
For small deviation in electron density from the original calculated model one might calculate $\delta B_{L}(q,q')$ to obtain directly the deviation in electron density using a similar method as introduced by Pande \it{et. al.} \cite{Pande2014}.
\\

\section*{Acknowledgement}

This work is partly supported by the National Science Foundation  Science and Technology Center (STC-1231306). I also acknowledge the UWM High Performance Computing Center (HPC) for the use of the Avi cluster.


\begin{thebibliography}{}
\expandafter\ifx\csname natexlab\endcsname\relax\def\natexlab#1{#1}\fi
\expandafter\ifx\csname bibnamefont\endcsname\relax
  \def\bibnamefont#1{#1}\fi
\expandafter\ifx\csname bibfnamefont\endcsname\relax
  \def\bibfnamefont#1{#1}\fi
\expandafter\ifx\csname citenamefont\endcsname\relax
  \def\citenamefont#1{#1}\fi
\expandafter\ifx\csname url\endcsname\relax
  \def\url#1{\texttt{#1}}\fi
\expandafter\ifx\csname urlprefix\endcsname\relax\def\urlprefix{URL }\fi
\providecommand{\bibinfo}[2]{#2}
\providecommand{\eprint}[2][]{\url{#2}}



\bibitem[{\citenamefont{Kam}(2009)\citenamefont{Z. Kam}}]{Kam1977}
  \bibinfo{author}{\bibfnamefont{Z.}~\bibnamefont{Kam}},
  \bibinfo{journal}{Macromolecules} \textbf{\bibinfo{volume}{10}},
  \bibinfo{pages}{927} (\bibinfo{year}{1977}).
  
  \bibitem[{\citenamefont{Elser}(2011)\citenamefont{V. Elser}}]{Elser2011}
  \bibinfo{author}{\bibfnamefont{V.}~\bibnamefont{Elser}},
  \bibinfo{journal}{New Journal of Physics} \textbf{\bibinfo{volume}{13}},
  \bibinfo{pages}{123014} (\bibinfo{year}{2011}).
  
   \bibitem[{\citenamefont{Namba et~al.}(2011)\citenamefont{K. Namba}}]{Namba1986}
  \bibinfo{author}{\bibfnamefont{K.}~\bibnamefont{Namba}} \bibnamefont{and}
  \bibinfo{author}{\bibfnamefont{G.} \bibnamefont{Stubbs}}, 
   \bibinfo{journal}{ Science} \textbf{\bibinfo{volume}{231}},
  \bibinfo{pages}{1401} (\bibinfo{year}{1986}).
  
  \bibitem[{\citenamefont{Watson  et~al.}(1953)\citenamefont{J. D. Watson}}]{Watson1953}
 \bibinfo{author}{\bibfnamefont{J. D.}~\bibnamefont{Watson}} \bibnamefont{and}
   \bibinfo{author}{\bibfnamefont{F. H. C.} \bibnamefont{Crick}},
   \bibinfo{journal}{ Nature(London)} \textbf{\bibinfo{volume}{171}},
  \bibinfo{pages}{737} (\bibinfo{year}{1953}).
  
  
    \bibitem[{\citenamefont{Saldin et~al.}(2009)\citenamefont{D K Saldin}}]{Saldin2009}
  \bibinfo{author}{\bibfnamefont{D. K.}~\bibnamefont{Saldin}},
  \bibinfo{author}{\bibfnamefont{V. L.} \bibnamefont{Shneerson}},
\bibinfo{author}{\bibfnamefont{R.}~\bibnamefont{Fung}} \bibnamefont{and}
  \bibinfo{author}{\bibfnamefont{A.} \bibnamefont{Ourmazd}},
  \bibinfo{journal}{J. Phys.: Condens. Matter} \textbf{\bibinfo{volume}{21}},
  \bibinfo{pages}{134014} (\bibinfo{year}{2009}).
  
  \bibitem[{\citenamefont{Pande et~al.}(2014)\citenamefont{K. Pande}}]{Pande2014}
      \bibinfo{author}{\bibfnamefont{K.}~\bibnamefont{Pande}},
  \bibinfo{author}{\bibfnamefont{P.}~\bibnamefont{Schwander}},
  \bibinfo{author}{\bibfnamefont{M.}~\bibnamefont{Schmidt}} \bibnamefont{and}
\bibinfo{author}{\bibfnamefont{D. K.}~\bibnamefont{Saldin}}, 
    \bibinfo{journal}{Phil. Trans. R. Soc. B} \textbf{\bibinfo{volume}{369}},
  \bibinfo{pages}{20130332} (\bibinfo{year}{2014}).

  
   
  \bibitem[{\citenamefont{Poon et~al.}(2009)\citenamefont{H. -C. Poon}}]{Poon2013}
   \bibinfo{author}{\bibfnamefont{H. -C.} \bibnamefont{Poon}}, 
  \bibinfo{author}{\bibfnamefont{P.}~\bibnamefont{Schwander}},
  \bibinfo{author}{\bibfnamefont{M.} \bibnamefont{Uddin}} \bibnamefont{and}
\bibinfo{author}{\bibfnamefont{D. K.}~\bibnamefont{Saldin}}, 
    \bibinfo{journal}{Phys. Rev. Letter} \textbf{\bibinfo{volume}{110}},
  \bibinfo{pages}{265505} (\bibinfo{year}{2013}).


\bibitem[{\citenamefont{Saldin et~al.}(2011)\citenamefont{D. K. Saldin}}]{Saldin2011}
  \bibinfo{author}{\bibfnamefont{D. K.}~\bibnamefont{Saldin}},
  \bibinfo{author}{\bibfnamefont{H. -C.} \bibnamefont{Poon}}, 
    \bibinfo{author}{\bibfnamefont{P.}~\bibnamefont{Schwander}}, 
     \bibinfo{author}{\bibfnamefont{M.} \bibnamefont{Uddin}} \bibnamefont{and}
\bibinfo{author}{\bibfnamefont{M.}~\bibnamefont{Schmidt}}, 
    \bibinfo{journal}{Opt. Express} \textbf{\bibinfo{volume}{19}},
  \bibinfo{pages}{17318} (\bibinfo{year}{2011}).

  \bibitem[{\citenamefont{AKirian}(2012)\citenamefont{R. A. Kirian}}]{Kirian2012}
 \bibinfo{author}{\bibfnamefont{R.}~\bibnamefont{A. Kirian}},  
   \bibinfo{journal}{J. Phys. B: At. Mol. Opt. Phys.} \textbf{\bibinfo{volume}{45}},
  \bibinfo{pages}{223001} (\bibinfo{year}{2012}).

\bibitem[{\citenamefont{Klug  et~al.}(1958)\citenamefont{A. Klug}}]{Klug1999}
 \bibinfo{author}{\bibfnamefont{A.}~\bibnamefont{Klug}},
   \bibinfo{journal}{Phil. Trans. R. Soc. B} \textbf{\bibinfo{volume}{354}},
  \bibinfo{pages}{531} (\bibinfo{year}{1999}).

\bibitem[{\citenamefont{Millane}(1991)\citenamefont{R. P. Millane}}]{Millane1991}
 \bibinfo{author}{\bibfnamefont{R. P.}~\bibnamefont{Millane}},  
   \bibinfo{journal}{Acta Cryst.} \textbf{\bibinfo{volume}{A47}},
  \bibinfo{pages}{449} (\bibinfo{year}{1991}).

 \bibitem[{\citenamefont{Kam}(2009)\citenamefont{Z. Kam}}]{Kam1982}
  \bibinfo{author}{\bibfnamefont{Z.}~\bibnamefont{Kam}},
  \bibinfo{journal}{J. Theor. Biol} \textbf{\bibinfo{volume}{82}},
  \bibinfo{pages}{15} (\bibinfo{year}{1980}).
  
  \bibitem[{\citenamefont{Cochran et~al.}(2011)\citenamefont{W. Cochran}}]{Cochran1952}
  \bibinfo{author}{\bibfnamefont{W.}~\bibnamefont{Cochran}}, 
  \bibinfo{author}{\bibfnamefont{F.H.C.} \bibnamefont{Crick}} \bibnamefont{and}
  \bibinfo{author}{\bibfnamefont{V.}~\bibnamefont{Vand}}, 
   \bibinfo{journal}{ Acta Cryst} \textbf{\bibinfo{volume}{5}},
  \bibinfo{pages}{581} (\bibinfo{year}{1952}).
  
  
\bibitem[{\citenamefont{Klug  et~al.}(1958)\citenamefont{A. Klug}}]{Klug1958}
 \bibinfo{author}{\bibfnamefont{A.}~\bibnamefont{Klug}},
  \bibinfo{author}{\bibfnamefont{F. H. C.}~\bibnamefont{Crick}} \bibnamefont{and}
   \bibinfo{author}{\bibfnamefont{H. W.} \bibnamefont{Wyckoff}},
   \bibinfo{journal}{Acta Cryst.} \textbf{\bibinfo{volume}{11}},
  \bibinfo{pages}{199} (\bibinfo{year}{1958}).
  
   \bibitem[{\citenamefont{Saldin et~al.}(2011)\citenamefont{D. K. Saldin}}]{Saldin2010}
 \bibinfo{author}{\bibfnamefont{D. K.}~\bibnamefont{Saldin}},
   \bibinfo{author}{\bibfnamefont{V L} \bibnamefont{Shneerson}},
  \bibinfo{author}{\bibfnamefont{D.} \bibnamefont{Starodub}} \bibnamefont{and}
  \bibinfo{author}{\bibfnamefont{J. C. H.}~\bibnamefont{Spence}}, 
   \bibinfo{journal}{ Acta Cryst} \textbf{\bibinfo{volume}{A66}},
  \bibinfo{pages}{32} (\bibinfo{year}{2010}).
    
     
 \bibitem[{\citenamefont{Donatelli et~al.}(2015)\citenamefont{J F Donatelli}}]{Donatelli2015}
  \bibinfo{author}{\bibfnamefont{J. F.}~\bibnamefont{Donatelli}},
  \bibinfo{author}{\bibfnamefont{P. H.} \bibnamefont{Zwart}} \bibnamefont{and}
  \bibinfo{author}{\bibfnamefont{J. A.} \bibnamefont{Sethian}},
  \bibinfo{journal}{Proceedings of the National Academy of Sciences} \textbf{\bibinfo{volume}{112}},
  \bibinfo{pages}{10286} (\bibinfo{year}{2015}).
    
  \bibitem[{\citenamefont{Liu et~al.}(2012)\citenamefont{H Liu}}]{Liu2012}
  \bibinfo{author}{\bibfnamefont{H.}~\bibnamefont{Liu}},
  \bibinfo{author}{\bibfnamefont{B. K.} \bibnamefont{Poon}},
\bibinfo{author}{\bibfnamefont{A. J. E. M}~\bibnamefont{Janssen}} \bibnamefont{and}
  \bibinfo{author}{\bibfnamefont{P. H.} \bibnamefont{Zwart}},
  \bibinfo{journal}{Acta Cryst} \textbf{\bibinfo{volume}{A68}},
  \bibinfo{pages}{561} (\bibinfo{year}{2012}).
  
   \bibitem[{\citenamefont{Liu et~al.}(2012)\citenamefont{H Liu}}]{Liu2013}
  \bibinfo{author}{\bibfnamefont{H.}~\bibnamefont{Liu}},
  \bibinfo{author}{\bibfnamefont{B. K.} \bibnamefont{Poon}},
   \bibinfo{author}{\bibfnamefont{D. K.}~\bibnamefont{Saldin}},
   \bibinfo{author}{\bibfnamefont{J. C. H.}~\bibnamefont{Spence}} \bibnamefont{and}
  \bibinfo{author}{\bibfnamefont{P. H.} \bibnamefont{Zwart}},
  \bibinfo{journal}{Acta Cryst} \textbf{\bibinfo{volume}{A69}},
  \bibinfo{pages}{365} (\bibinfo{year}{2013}).
  
   \bibitem[{\citenamefont{Oszlanyi  et~al.}(2011)\citenamefont{G. Oszlanyi}}]{Oszlanyi2004}
 \bibinfo{author}{\bibfnamefont{G.}~\bibnamefont{Oszlanyi}} \bibnamefont{and}
   \bibinfo{author}{\bibfnamefont{A.} \bibnamefont{Suto}},
   \bibinfo{journal}{ Acta Cryst.} \textbf{\bibinfo{volume}{A60}},
  \bibinfo{pages}{134} (\bibinfo{year}{2004}).
  
   \bibitem[{\citenamefont{Oszlanyi  et~al.}(2011)\citenamefont{G. Oszlanyi}}]{Oszlanyi2005}
 \bibinfo{author}{\bibfnamefont{G.}~\bibnamefont{Oszlanyi}} \bibnamefont{and}
   \bibinfo{author}{\bibfnamefont{A.} \bibnamefont{Suto}},
   \bibinfo{journal}{ Acta Cryst.} \textbf{\bibinfo{volume}{A61}},
  \bibinfo{pages}{147} (\bibinfo{year}{2005}).
  
 \bibitem[{\citenamefont{HHayes}(2012)\citenamefont{M. H. Hayes}}]{Hayes1982}
 \bibinfo{author}{\bibfnamefont{M. H.}~\bibnamefont{M. H. Hayes}},  
   \bibinfo{journal}{IFFF Trans Acoust Speech Signal Process} \textbf{\bibinfo{volume}{30(2)}},
  \bibinfo{pages}{140} (\bibinfo{year}{1982}).
  


\end{thebibliography}
\end{document}